\documentstyle[12pt,aps]{revtex}
\begin{document}
\def\gsimeq{\; \raisebox{-0.4ex}{\tiny$\stackrel
{{\textstyle>}}{\sim}$}\;}
\def\lsimeq{\; \raisebox{-0.4ex}{\tiny$\stackrel
{{\textstyle<}}{\sim}$}\;}
\title{Tri-axial Octupole Deformations and Shell Structure}
\author{\sc W.D. Heiss$^1$, R.G. Nazmitdinov$^{1,2}$ and R.A. Lynch$^1$}
\address{$^1${\sl Centre for Nonlinear Studies and Department of Physics\\
University of the Witwatersrand,
PO Wits 2050, Johannesburg, South Africa\\
$^2$ Bogoliubov Laboratory of Theoretical Physics\\
Joint Institute for Nuclear Research, 141980 Dubna, Russia}}
\maketitle
\vspace{0.3cm}
\begin{abstract}
Manifestations of pronounced shell effects are discovered when adding
nonaxial octupole deformations to a harmonic
oscillator model. The degeneracies of the quantum spectra
are in a good agreement with the corresponding main periodic orbits and
winding number ratios which are found by classical analysis.

\end{abstract}

\vspace{0.3cm}

The remarkable regularity of rotational spectra of
superdeformed nuclei has prompted many investigations into
the contribution of higher multipoles on the formation
of shell structure \cite{Ab}. Related
questions have arisen for other mesoscopic systems.
In particular, it appears that octupole deformation
has the same importance for two-dimensional systems like
quantum dots and surface clusters \cite{Man} as
for three-dimensional systems like nuclei \cite{Naz} and
metallic clusters \cite{Frau}.
Semi-classical analysis based on periodic orbit theory
 \cite{BB,BM,Br} provides substantial insight
into  the role of the octupole deformation
for axially symmetric systems \cite{hena,Ar,Cr}.
Axially symmetric octupole deformation which leads to soft chaos in
the classical case produces short periodic orbits
at particular parameter strengths, and, correspondingly, pronounced shell
effects arise in the quantum spectrum \cite{hena,Ar}.

Conservation of angular momentum may
increase the regular region for a non-integrable problem
with axial symmetry (see for example \cite{HR}).
The situation becomes more complicated for nonaxial systems with
three degrees of freedom, since angular momentum is no longer a constant
of motion and the classical dynamics may lead to a stronger degree of
chaos \cite{Gut}. Inclusion of exotic, i.e. nonaxial, octupole deformations
renders the finding of pronounced shell effects rather difficult.
Results based on the term $Y_{3\pm 1}$, which was suggested by
Mottelson \cite{Mot}
and studied in \cite{Chas}, have been questioned in \cite{Sk}.
Other attempts which incorporate nonaxial octupole deformation
start with axially symmetric potentials; they have found indications
of shell effects using $Y_{3\mu}$ deformations mainly with $\mu=0,2$
\cite{Sk,Ham}. The increasing accuracy of measurements
of nuclear spectra, due to the new generation of
detectors, gives substantial indications of
strong octupole correlations \cite{Naz,Hw}. This calls
for a thorough analysis of nonaxial octupole deformations.
A similar question about
shell effects in electronic structures and their connection
to octupole deformations arises from
the {\it ab initio} calculations of the melting transitions
in small alkali clusters \cite{Ryt}. In this Letter we demonstrate
the existence of strong shell effects which arise in the
tri-axial harmonic oscillator combined with nonaxial octupole deformations.
The model may serve as a simple and transparent study of the effective
mean field for mesoscopic systems like nuclei and
metallic clusters. In addition, features characteristic of
realistic potentials, i.e. a coexistence of regular and chaotic
dynamics and the consequences for quantum mechanics, are addressed.

Similar to the procedure pursued in \cite{hena} we are guided by the study
of the classical motion in obtaining the quantum mechanical results. The
single--particle Hamiltonian considered reads
\begin{equation}
H={(\vec p)^2\over 2m}+{m\omega ^2\over 2}\left[({ x \over a})^2
+({y \over b})^2+({z \over c})^2
+r^2\sum_{\mu} \lambda _{3\mu }Y_{3\mu }\right]  \label{ham}
\end{equation}
where the $Y_{\lambda \mu }$ are the usual spherical harmonics. To
maintain time reversal invariance, only the combinations
$Y_{3\mu}\pm Y_{3-\mu}$ are considered, with a factor $i$ where
appropriate. We take into account only one of the deformation
$\mu=0,1,2,3$ at a time. For convenience, we express all
quantities below in units of $\omega \equiv \omega_x$, i.e. $a=1$.
If any of the parameters $\lambda _{3\mu}$ are nonzero, we are
faced with a non-integrable system. In fact, the problem gives
rise to chaotic motion even at relatively small values of the
octupole parameters. The parameters have to be limited by their
respective critical values $\lambda_{3\mu }^{crit}$ at which the
potential no longer binds. The critical values depend on the
parameters $a$, $b$, $c$ which can be expressed through standard
quadrupole deformations $\epsilon _{2\mu }$ and, if more than one
$\lambda _{3\mu}$ is considered, critical surfaces are obtained.
It is obvious that a search for shell structure for the
corresponding quantum mechanical problem becomes meaningless above
the critical values as then the quantum mechanical spectrum
obtained by matrix diagonalisation does not relate to the
corresponding classical Hamiltonian.

The quadrupole shapes as determined by the parameters
$a$, $b$, $c$ are illustrated by the hexagonal
figure given in Fig.1. The three
axes denoting axially symmetric prolate (oblate) shapes differ by an
appropriate permutation of the coordinates $x-y-z$. For physical
consideration it is therefore sufficient to consider just one sector if
only quadrupole deformation is being studied \cite{BM}.
However, the addition of an octupole term defines an orientation,
since it refers to a specific
$z$-axis. Therefore, when adding, say, the term $r^2\lambda _{30}Y_{30}$,
the physical situation is different for two oblate cases, for example,
$obl/x$ and $obl/z$ in Fig.1.
Note that the latter case would preserve axial symmetry,
thus making it effectively a two degrees of freedom system.
But the former case is now a genuine three degrees of freedom system,
since the symmetry axis of the quadrupole shape (the $x$-axis)
is different from the symmetry axis of $r^2\lambda _{30}Y_{30}$
(the $z$-axis).

In the same vein, the addition of a term
$r^2\lambda _{3\mu }Y_{3\pm \mu}$ to any
of the axially symmetric quadrupole shapes, gives rise to a three degrees
of freedom system for $\mu \ne 0$.

The effect of $r^2Y_{30}$ upon the axes $pro/z$ and $obl/z$ has been
dealt with in \cite{hena}. We here begin with  the combination
$r^2(Y_{33}- Y_{3-3})$. It appears from numerical analysis of
the classical equations of motion (see below) that the
most regular motion occurs in the vicinity of the $pro/x$ axis.
The procedure to approximate the non-integrable classical system
is the 'removal of resonances' method \cite{LL}.
To lowest order it consists of
averaging the Hamilton function over the fastest angle of the unperturbed
motion (all $\lambda_{3\mu} =0$)
after rewriting the momenta and coordinates in terms of action--angle
variables, {\it viz.}
\begin{eqnarray}
& q_i=\sqrt{{2J_i\over m\omega_i}} \sin\theta _i , \quad
p_i=\sqrt{2J_i m \omega_i}\cos\theta _i, \nonumber \\
& \theta _i=\omega_it , \quad i=x,y,z. \nonumber
\end{eqnarray}
On the axis $pro/x$ one would therefore average over $\theta _z$
or $\theta _y$ and above it over $\theta _z$, since there $\omega
_z>\omega_y>\omega_x$ ($a>b>c$). Based on the observation that
$r^2(Y_{33}-Y_{3-3})$ is proportional to $x(x^2-3y^2)/r$, we
expect the $z$-motion to be weakly affected by this term. In the
vicinity of the $pro/x$ axis an averaging over $\theta _z$ is thus
indicated. Moreover, for the reasons just given, an averaging over
$\theta _z$, which yields an effective potential in the
$x-y$-coordinates, is expected to make little difference from an
effective potential obtained by simply setting $z=0$. This
expectation is convincingly confirmed by numerical tests as long
as $\omega _z\ge \omega_y$. In other words, for $\omega_z\ge
\omega_y > \omega_x$, the motion effectively decouples into an
unperturbed motion in the $z$-coordinate (governed by the
potential $mz^2\omega ^2/(2c^2)$) and the two degrees of freedom
motion in the $x-y$-plane. Averaging now over the fast angle
$\theta _y$ yields the unperturbed motion in the $y$-coordinate
(governed by $my^2\omega ^2/(2b^2)$) and the effective potential
for $x$ which reads
\begin{eqnarray}
& U_{{\rm eff}}(x)={m\omega ^2\over 2}\bigl[x^2+ \nonumber \\
& \lambda _{33}
{{\rm sign}(x)\over \pi}\bigl(2x^2K(-{\xi_y^2\over x^2})-
3\pi \xi_y^2{_2}F_{1}({1\over2},{3\over 2},2;-{\xi _y^2\over x^2})\bigr)\bigr]
\label{ux}
\end{eqnarray}
where $K$ and ${_2}F_{1}$ denote the first elliptic integral and
the hypergeometric function, respectively, and $\xi _y^2=2E_y/(m
\omega _y^2)$ where $\omega_y= \omega /b$. The approximation used
here assumes that $E_y$, the energy residing in the $y$-motion, is
constant (and therefore also $E_x$); note that the effective
potential $U_{{\rm eff}}(x)$ depends on $E_y$. A numerical
comparison between the true three dimensional motion and the
approximate decoupled motion nicely confirms the validity of the
approximation for $\lambda _{33}\lsimeq \lambda_{{\rm crit}}/2$.
The crucial test which is relevant for the corresponding quantum
mechanical case is the comparison of the winding number ratios
$\omega _x^{{\rm eff}}/\omega _y$ and $\omega _x^{{\rm
eff}}/\omega _z$. Moreover, these ratios are virtually independent
of $E_y$ for $E_y$ less than 60\% of its maximal value $E_{{\rm
tot}}$. As a result, we may evaluate analytically $\omega _x^{{\rm
eff}}$ by choosing $E_y=0$ in Eq.(\ref{ux}) and obtain
\begin{equation}
\label{rat}
\omega _x^{{\rm eff}}:\omega_y:\omega_z = \frac{2\sqrt{1-\lambda_{33}^2}}
{\sqrt{1-\lambda_{33}}+\sqrt{1+\lambda _{33}}}:{1\over b}:
{1\over c}
\end{equation}

By appropriately tuning the parameters $b,c$ and $\lambda _{33}$,
we can obtain simple ratios for the winding numbers. They
determine short periodic orbits which are expected to occupy a
major part of a phase space.

For a two--dimensional problem the Poincar\'e surface of section
can be used for the estimation of the percentage of phase space
occupied by periodic orbits. In the present situation, the five
dimensional phase space renders an understanding of the underlying
structure rather difficult. The technique utilized here is
essentially a frequency analysis. It is in principle impossible to
determine whether a trajectory is quasi-periodic or chaotic merely
by looking at the frequency spectra \cite{DB}.  A pragmatic
approach is adopted here in that an initial condition is deemed to
yield a chaotic orbit if the associated frequency spectra have
sufficiently many discernible peaks. We call an orbit chaotic if
any one of the frequency spectra has more than six peaks with
intensities greater than 1\% of the maximum intensity. These
arbitrary choices proved satisfactory for our purpose. If the
orbit is quasi-periodic, then the most significant frequency peaks
are compared, and the approximate winding number and period
obtained. Repeating this procedure many times using different
initial conditions in phase space yields a Monte-Carlo type
estimate of the portions of phase space characterized by the
various different frequency ratios. If a particular simple ratio
dominates, as exemplified below, then we expect specific
signatures in the quantum spectrum as shell structure. Examples
are illustrated in Figs.2--4. We have chosen the numbers
$a:b:c=1:0.5577:0.5577,1:0.3718:0.5577$ to obtain the ratios
$1:2:2,1:3:2$ from Eq.(\ref{rat}), respectively. This ratio is
sufficiently simple to make an easy comparison with the quantum
results.

The adiabatic approach predicts orbits with the frequency ratios
$1:2:2$ and $1:3:2$ at $\lambda_{33}=0.5 \lambda_{crit}$. The
corresponding spectra are displayed in Fig.\ref{fig2}. The
classical frequency analysis (CFA) of the exact orbits shows a
peak at $\lambda_{33}\approx 0.55\lambda_{33}^{crit}$ for the
ratios $1:2:2$,$1:3:2$, and the quantum shell structure occurring
at $\lambda_{33}\approx 0.5\lambda_{33}^{crit}$ has the correct
degeneracy pattern for about the first hundred levels.

As a quantitative measure for shell structure we use
the Strutinsky-type analysis introduced in \cite{hena}. From the quantity
$\Delta E(\lambda, N) = \delta E(\lambda, N+1)+\delta E(\lambda, N-1)-
2\delta E(\lambda, N)$, where $\delta E$ is the fluctuating part of the total
energy, we obtain the precise location of the magic numbers (see Fig.4).
Similarly, the whole discussion can also be applied to the
axis $pro/y$ in Fig.1 by using the combination
$r^2(Y_{33}+Y_{3-3})\sim y(y^2-3 x^2)/r$ instead.
It appears that, due to the weak $z$-dependence of
the combinations $r^2(Y_{33}\pm Y_{3-3})$,
mainly the $x-y$ profile of the unperturbed harmonic oscillator is
important. The stronger the deformation in the $x-y$-plane,
i.e. the further away from the $z$-axial symmetry line ($pro/z$ and $obl/z$),
the better the adiabatic approximation and shell structure becomes.
In contrast, along this horizontal line either
$Y_{3\pm 3}$  combination acts upon a circular potential in the $x-y$-plane
and quickly introduces chaos.

The CFA, applied to the combination
$r^2(Y_{31}-Y_{3-1}) \sim x(4z^2-x^2-y^2)$,
reveals that the $y$ motion is weakly affected
for $\lambda_{31}\lsimeq 0.6\lambda_{31}^{crit}$ in
the vicinity of the $pro/x$ axis. Applying the analysis described
above, we found that this term leads to shell effects
similar to those of $r^2(Y_{33}- Y_{3-3})$.
In fact, Eq.(\ref{rat}) holds  for this combination as well.
The plus combination $r^2(Y_{31}+Y_{3-1})$ is simply the minus
combination under the interchange of $x$ and $y$, and thus will produce
the same effects near the region $pro/y$.
It is important to note that the addition of a term
$\lambda _{30}r^2 Y_{30}$ to either situation, $pro/x$ or $pro/y$, leads to an
onset of chaos for rather small values of $\lambda _{30}$, and accordingly
the quantum spectrum does not exhibit shell structure.

The two combinations
$r^2(Y_{3\mu }\pm Y_{3-\mu })$ produce the same potential shape
 for the spherical case \cite{Ham}. However, according to the
analysis above, the
plus and minus combinations have different effects
for different sectors of the hexagonal figure (Fig.1).
The CFA shows that the adding of the octupole term
$\lambda _{32}r^2(Y_{32}+Y_{3-2})$ gives rise to chaotic motion for
comparatively small coupling values. In contrast, the term
$\lambda _{32}r^2 (Y_{32}-Y_{3-2})$ has less impact on the unperturbed
motion, chaotic motion only becoming discernible for $\lambda _{32}\gsimeq
0.5\lambda _{{\rm crit}}$.
Since $(Y_{32}-Y_{3-2})$ is symmetric with respect to an
interchange of $x$ and $y$, this result applies to the region above and
below the axis $pro/z$ including the axes $obl/x$ and $obl/y$. The quantum
mechanical results are in accordance with the classical findings: the plain
quadrupole spectrum changes weakly over a considerable range of
$\lambda _{32}$ when adding the term $r^2 (Y_{32}-Y_{3-2})$ while the order
soon decays when the plus combination is switched on.

The cases considered here represent
novel examples of {\it three--dimensional} non-integrable systems,
which can be well-approximated by integrable ones.
The results of the current literature are limited
primarily to axially symmetric non-integrable systems \cite{Br,Gut}.
Special values of parameters, found with
the 'removal of resonances' method, produce
potentials conducive to regular classical motion in much of
phase space. The various octupole combinations may have different
effects on the generation of shell structure, depending on
where the unperturbed potential lies in the hexagonal figure (Fig.1).
The effect of $r^2 Y_{30}$ upon the axis $pro/z$ is similar to
that of $r^2(Y_{33}\pm Y_{3-3})$ and $r^2(Y_{31}\pm Y_{3-1})$
upon the axis $pro/x$ or $pro/y$.
In contrast, the terms $r^2 (Y_{32}\pm Y_{3-2})$ do
not support shell structure.
In this context we mention that the special combination
$r^2\tilde \lambda (Y_{30}+ 3(Y_{32}+Y_{3-2}))$, when added to $pro/z$, is,
after suitable permutation of the coordinates, identically
equivalent to the adding of $r^2\lambda_{33}(Y_{33}- Y_{3-3})$ to $pro/x$.
Thus, we have generalized our
previous result \cite{hena} to the domain of triaxiality in that
the combination of quadrupole and {\it nonaxial} octupole
deformations has been shown to lead to shell effects equivalent to those
from a plain quadrupole deformed potential, at least for the first one
hundred levels. More comprehensive details of the classical frequency
analysis, the adiabatic approach and the quantum mechanical analysis will
be presented in a forthcoming paper.
Finally, we note that negative parity states
observed in rare-earth nuclei with neutron number $N\sim 92$,
 which become yrast at high spins,
need an unexpectedly strong degree of triaxiality
(see \cite{tr2}) when described in terms of
quadrupole and hexadecapole deformations only. We suggest that
inclusion of an octupole deformation $r^2(Y_{33}-Y_{3-3})$
or a banana-type octupole deformation $r^2(Y_{31}-Y_{3-1})$ which effectively
gives rise to shell effects of a tri-axial oscillator, could yield
a more natural explanation of these phenomena.

R.G.N. acknowledges financial support from the Foundation for Research
Development of South Africa which was provided under the auspices of the
Russian/South African Agreement on Science and Technology.

\newpage

\centerline{Figure Captions}

{\bf Fig.1}
Shapes in the $a$, $b$, $c$ plane. Spherical symmetry $(a=b=c)$
is at the center while axially
symmetric prolate and oblate shapes are obtained along the various axes.
A genuine tri-axial quadrupole deformation ($a\neq b \neq c$)
occurs between the axes.

\vspace{1cm}

{\bf Fig.2} Two quantum spectra: (a) for the parameters
$b=c=0.5577$; (b) for the parameters $b=0.3718$, $c=0.5577$.
Energies are given in units of $\hbar \omega $.

\vspace{1cm}

{\bf Fig.3} Classical phase space occupation for the frequency
ratios 1.1:3:2  and 1:3:2. The 1:3:2 peak at
$\lambda_{33}/\lambda_{33}^{crit}\approx 0.55$ leads to the shell
structure displayed in Fig.2b.

\vspace{1cm}

{\bf Fig.4} Magic numbers calculated at values of parameters as for
Fig.2. The magic numbers corresponding to a pure quadrupole
deformed harmonic oscillator are indicated in (a) for the ratio
$1:2:2$ and in (b) for the ratio $1:3:2$.

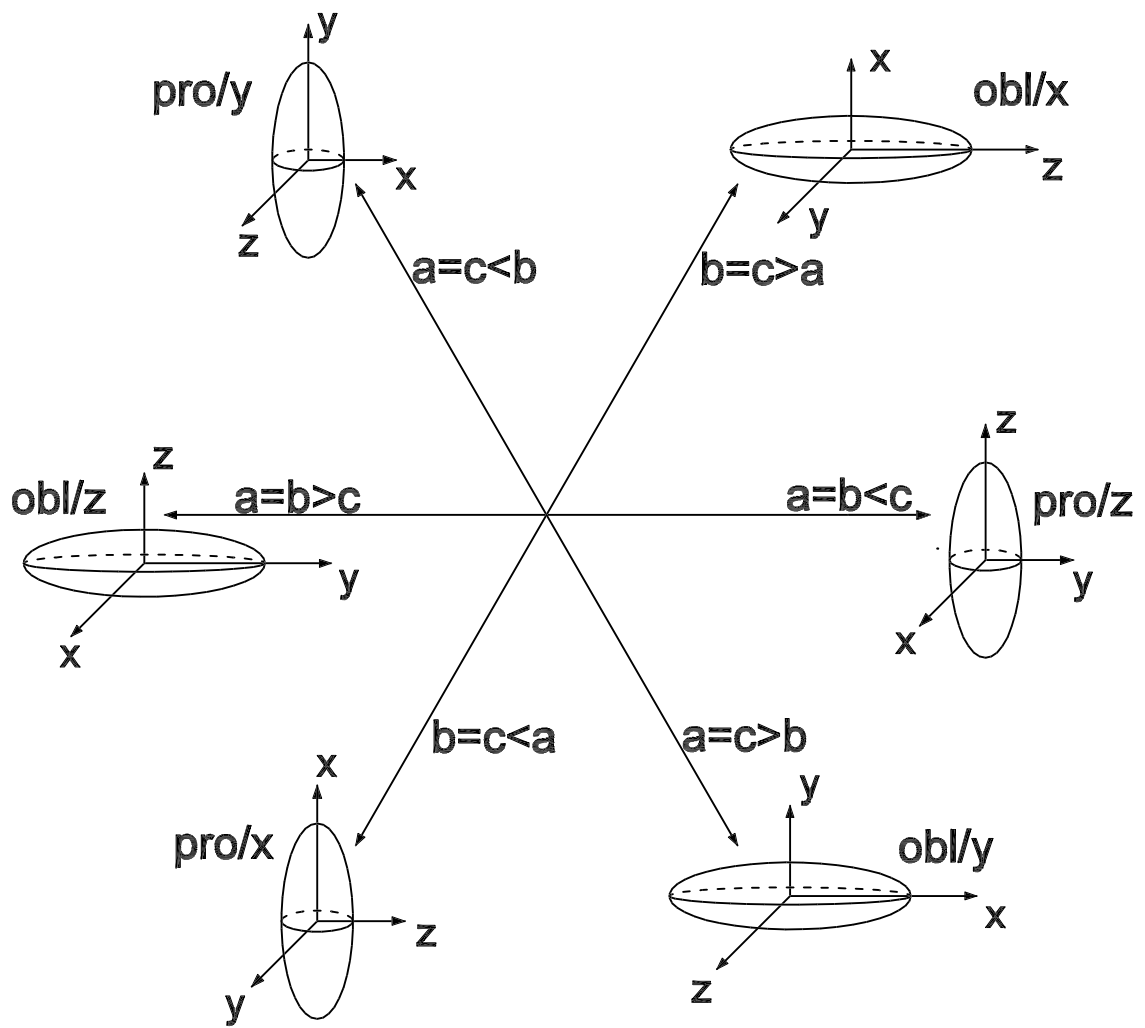
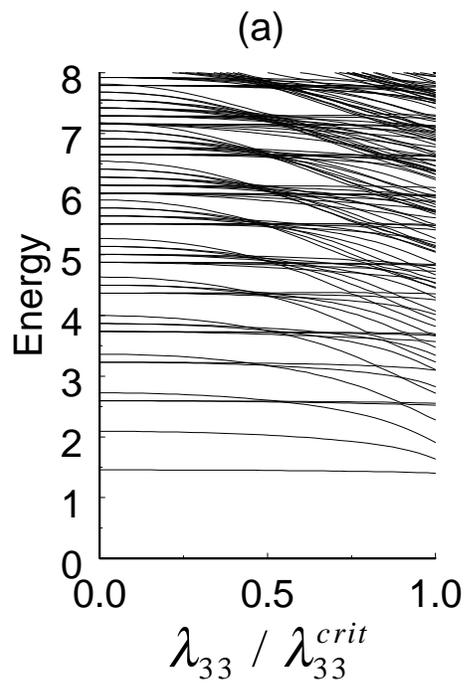
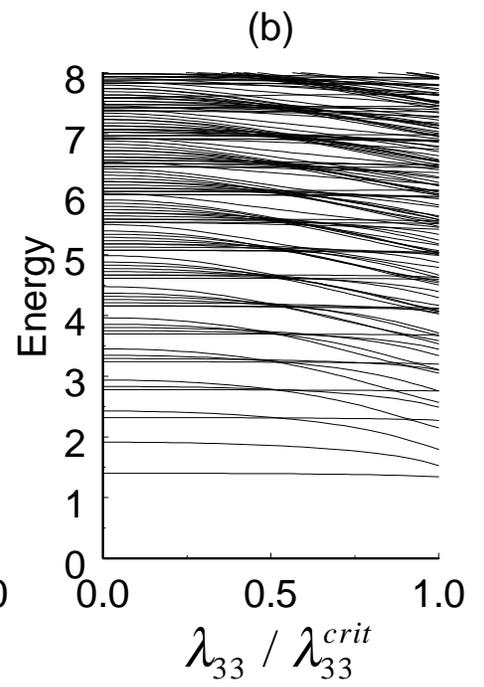
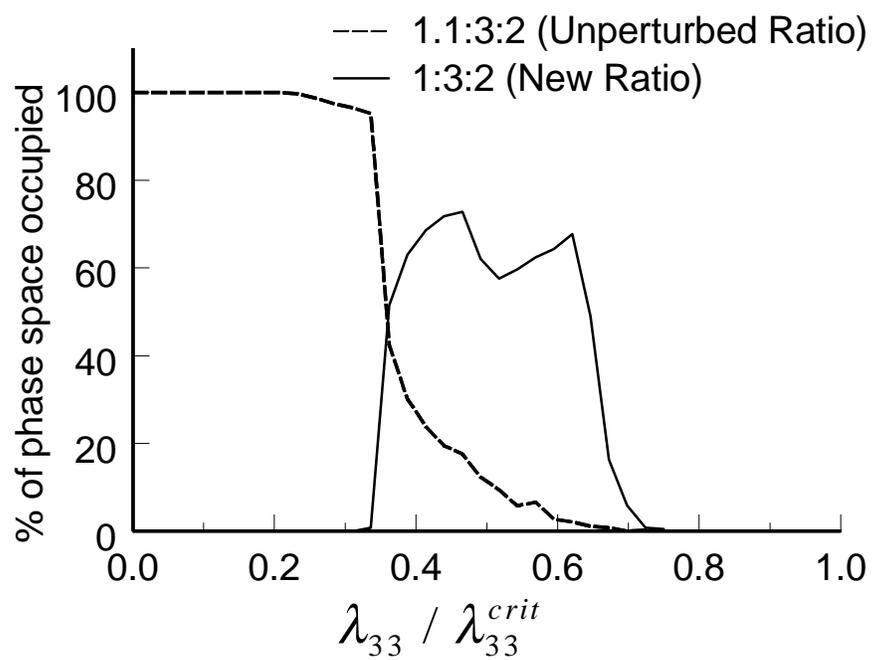
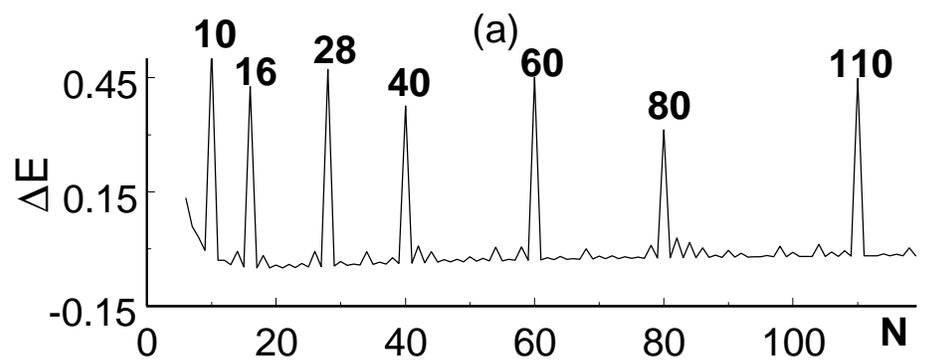
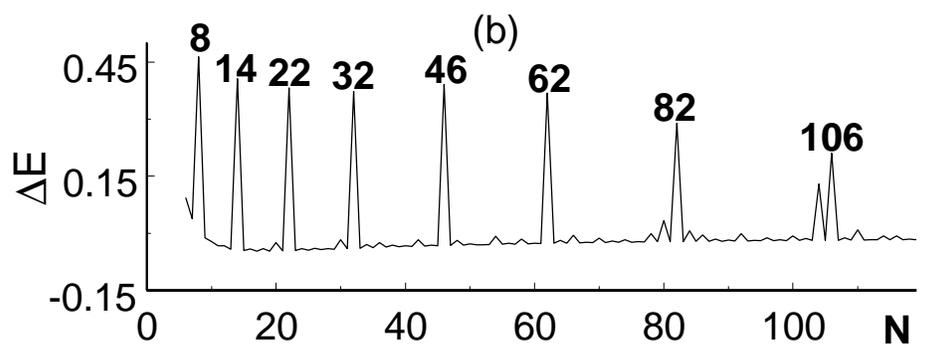
\end{document}